# HeatFlicker: A Virtual Campfire System Utilizing Flickering Thermal Illusions by Asymmetric Vibrations

Takato Ito[1], Takeshi Tanabe[2], Shoichi Hasegawa[3], Naoto Ienaga[4], and Yoshihiro Kuroda[4]

[1] *Graduate School of Science and Technology, University of Tsukuba, Ibaraki, Japan*

[2] *Human Informatics and Interaction Research Institute, National Institute of Advanced Industrial Science and Technology (AIST), Ibaraki, Japan*

[3] *Laboratory for Future Interdisciplinary Research of Science and Technology, Tokyo Institute of Technology, Tokyo, Japan*

[4] *Institute of Systems and Information Engineering, University of Tsukuba, Ibaraki, Japan*

(Email: ito@le.iit.tsukuba.ac.jp)

**Abstract ---** In recent years, thermal feedback has emerged as a significant sensory modality in virtual reality. However, the concept of conveying the sensation of thermal movement remains largely unexplored. We propose HeatFlicker, a virtual campfire device that recreates the flickering of fire by using a thermal illusion of moving heat identified in preliminary experiments. This device creates the illusion of heat moving from a fixed heat source. In our demonstration, we provide a novel thermal experience by simulating the flickering of a real fire.

## 1 INTRODUCTION

In recent years, research on non-contact thermal sensation feedback has been actively conducted to enhance the realism of virtual reality (VR). However, there has been limited discussion on the representation of moving thermal sensations within VR spaces. Physical methods for transferring heat often result in larger devices and impose mechanical limitations, such as restrictions on the locations and areas where thermal sensations can be presented. Therefore, approaches utilizing illusions to present the sensation of thermal movement without the actual physical transfer of heat are being studied. Thermocaress combined haptic feedback through pneumatic actuators and thermal feedback via circulating water, enabling the induction of a stroking thermal sensation using thermal referral [1]. Also, a method using apparent motion induced by thermal stimulation can be considered [2]. However, it requires several seconds of thermal stimulation before the illusion is induced, and the sensation of thermal movement becomes intermittent, potentially causing discrepancies with the intended experience. Moreover, traditional methods were contact-based, limiting the range of illusion induction and potentially influencing perception due to the tactile sensation. These sensory and temporal discrepancies and expression limitations can reduce the realism of VR experiences.

Therefore, inducing an illusion of moving thermal sensations with non-contact or minimally contact-based devices could flexibly adapt to changes in the presentation range and reduce discrepancies with the intended sensations. This approach is expected to achieve a more accurate and flexible representation of moving thermal sensations in VR.

In particular, expressing a small and rapid thermal movement, such as the sensation of flickering flames, often results in ambiguities and the required equipment tends to become larger relative to the presentation area when using conventional methods. "Flickering flames" refers to flames that sway or move unsteadily.

Green observed a phenomenon where the thermal sensation shifts to the location where the tactile stimulus is presented when tactile and thermal stimuli are simultaneously presented at different locations, and defined this phenomenon as thermal referral [3]. Considering this, we speculated that there might be an illusory relationship between vibratory and thermal sensations. We focused on asymmetric acceleration vibrations that can induce the pulling illusion using a small vibrator [4].

In this study, to avoid interference with the illusion due to the contact sensation of thermal elements, we employed non-contact thermal stimulation. Specifically, considering the ease of output adjustment via pulse width modulations, we used visible light emitting diodes (LED) for non-contact thermal stimulation [5].

In the preliminary experiment, which involved three participants including myself, the combination of non-contact thermal stimulation around the hand and asymmetric vibrations delivered to a vibrator held at the fingertips resulted in a sensation where the heat seemed to shift in synchrony with the pulling illusion.

Although there are previous studies, such as the system developed by Singhal et al. that utilizes an ultrasonic haptic display and thermal airflow to present the thermal sensations of embers from a campfire [6], there is currently no system designed to replicate the unique, unstable flickering thermal sensations of fire through illusions in virtual reality environments.

This study aims to provide a novel sensory experience where heat from a stationary heat source feels as though it is flickering by utilizing pulling illusions by asymmetric vibrations.

## 2 HEATFLICKER DEVICE

Fig.1 shows the configuration of the proposed thermal movement illusion device called HeatFlicker. HeatFlicker consists of a thermal display part with an integrated speaker and a device cover with a liquid crystal display (LCD) panel on its top. The thermal display part features multiple LED chips arranged on slanted sides. Additionally, a fan is installed at the bottom to dissipate the heat generated by the LEDs.

The front of the device cover is equipped with a curtain that blocks the light from the LEDs and allows the user to insert both arms. An LCD panel is mounted on the top, with a camera positioned behind it to capture the interior of the device cover from above.

Furthermore, HeatFlicker employs a small vibrator that can be grasped with the fingertips. To induce the pulling illusions, asymmetric vibration stimuli developed in our previous study [7] were presented by the vibrator.

Additionally, Fig.2 shows the appearance of the prototype HeatFlicker device. The image was taken during the development phase, and actual demonstration may have some modifications to the device.

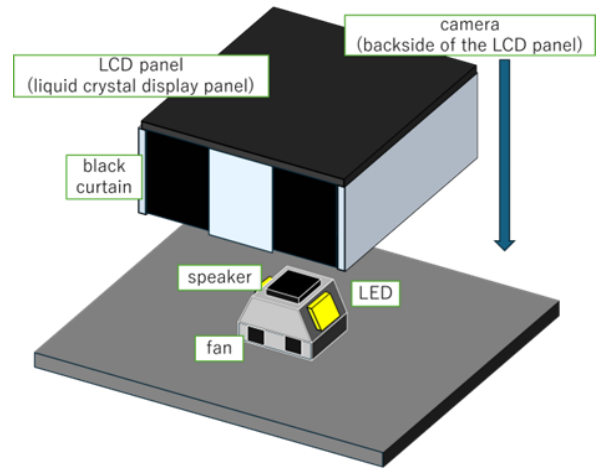

Fig.1 HeatFlicker Design: HeatFlicker consists of a thermal display component by LEDs and a device cover with an LCD panel.

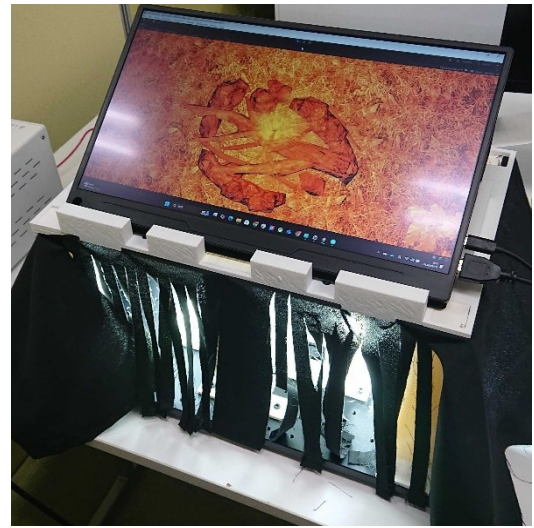

Fig.2 Appearance of the Prototype HeatFlicker: Actual demonstration may have some modifications to the device.

## 3 DEMONSTRATION

Fig.3 shows the demonstration setup. The users insert their arms into the curtain while grasping the vibrator between their thumbs and index fingers. HeatFlicker recognizes the user's hands from the internal camera footage and synchronizes the movements of virtual hands with the user's actions. The speaker in the device plays a crackling sound, like the sound of a campfire. Asymmetric vibrations are applied to the vibrator, periodically switching directions to create a pulling illusion, thereby inducing the illusion of thermal movement.

In this demonstration, we provide a novel experience of the flickering heat from a campfire in a VR environment.

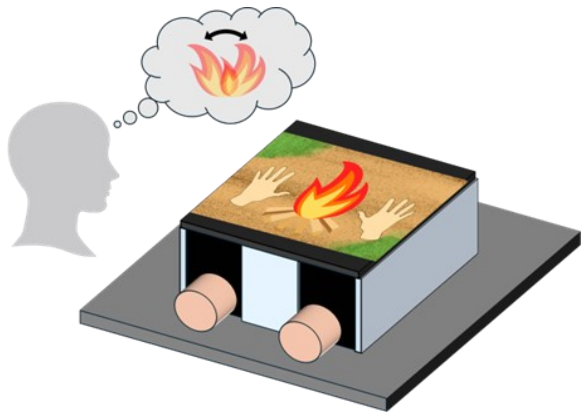

Fig.3  HeatFlicker Demonstration Image: HeatFlicker provides a novel experience of the flickering heat from a virtual campfire.

## 4  CONCLUSION

In this study, we present a demonstration system that allows users to experience the sensation of a flickering campfire by utilizing the flickering thermal illusion. However, for this thermal illusion, quantitative evaluations, such as the relationship between the direction of the pulling illusion and the perceived direction of thermal movement, as well as the intensity of the flickering thermal illusion, have not yet been conducted. Therefore, further research is required to explore the characteristics of this illusion.

### ACKNOWLEDGEMENT

This research was supported by JSPS KAKENHI Grant Numbers JP21H03474, JP24K02969, and JP24K22316.